\documentclass[aps,prb,showpacs,twocolumn,amssymb]{revtex4}

\usepackage{graphicx}
\usepackage{amsmath}

\input epsf

\def\prb{Phys. Rev. B }
\def\prl{Phys. Rev. Lett. }

\def\be{\begin{equation}}
\def\ee{\end{equation}}
\def\ba{\begin{eqnarray}}
\def\ea{\end{eqnarray}}

\def\124{YBa$_2$Cu$_4$O$_8$ }

\def\BSCCO{Bi$_2$Sr$_2$CaCu$_2$O$_{8+\delta}$}
\def\C60{A$_x$C$_{60}$ }
\def\LNSCOsc{La$_{1.6-x}$Nd$_{0.4}$Sr$_{x}$CuO$_{4}$ }

\def\LBCO{La$_{2-x}$Ba$_{x}$CuO$_4$}

\begin{document}

\title{Effects of geometrical fluctuations on the transition temperature of disordered
quasi-one-dimensional superconductors}

\author{Dror~Orgad}

\affiliation{Racah Institute of Physics, The Hebrew University,
Jerusalem 91904, Israel}

\date{\today}

\begin{abstract}

We study the superconducting transition temperature, $T_c$, of an array of fluctuating,
maximally gapped interacting ladders, embedded in a disordered plane.
Renormalization group analysis indicates that geometrical fluctuations mitigate
the suppression of the intra-ladder pairing correlations by the disorder. In addition, the
fluctuations enhance the Josephson tunneling between the ladders. Both effects lead to an
increase of $T_c$ in the Josephson coupled array. These findings may be relevant to the
understanding of the 1/8 anomaly in the lanthanum based high-temperature superconductors.

\end{abstract}


\maketitle

\section{Introduction}

Both perturbative renormalization group (RG) studies \cite{RGladders} and strong coupling
numerical calculations \cite{num-ladders} have demonstrated that doped $t-J$ and Hubbard
ladders typically exhibit a maximally gapped phase, with a single remaining gapless charge
mode protected by symmetry \cite{manycomp}. Such a phase possesses substantial
superconducting pairing correlations with approximate $d$-wave symmetry. By coupling
many ladders together via Josephson tunneling, the phases of the different Cooper pairs
may establish coherence, in which case global superconductivity ensues. It has been
suggested that the underlying mechanism of superconductivity in the cuprates is
essentially the one described here for the quasi-one-dimensional system
\cite{ourreview,steve-RMP}.

Previous studies have shown that Josephson coupled arrays of spin-gapped chains and
two-leg ladders are particularly prone to the deleterious effects of disorder on $T_c$
\cite{suzumura,orignac2l}. On the other hand, we have demonstrated that geometrical
fluctuations of a spin-gapless one-dimensional system tend to reduce the effects of
disorder \cite{fluctuating}. It is therefore interesting to try and marry these two
observations, and study the interplay between disorder and fluctuations in a
quasi-one-dimensional superconductor. Beyond its purely theoretical appeal this
question may also be relevant to a distinctive feature in the phenomenology of
the lanthanum based cuprates.

Among the high-temperature superconductors, the lanthanum compounds stand out as the
ones with the most extensive experimental evidence for the existence of
quasi-one-dimensional charge and spin stripe order, both in their "normal" and
superconducting states \cite{lanthanum,la-str-Tc}. Experiments have also found clear
correlation between the presence of robust static stripe order and the suppression of
superconductivity in these materials \cite{la-str-Tc}. For example, in \LNSCOsc the ordering
temperature, $T_m$, of the incommensurate magnetic structure reaches a maximum around
$x=1/8$, while the superconducting temperature, $T_c$, exhibits a substantial dip there.
This 1/8 anomaly was first identified in \LBCO \cite{LBCO1/8}, where it is especially
pronounced. In view of the theoretical insights outlined above, it is tempting to associate
the suppression of $T_c$ with enhanced disorder effects due to the slowing down of the stripe
fluctuations.

Motivated by such questions we study in this paper a model of a fluctuating disordered
array of maximally gapped $n$-leg ladders. We find that the fluctuations reduce the
disorder induced degradation of the pairing correlations on each one-dimensional system,
although they typically do not change the critical value of the Luttinger interaction
parameter, $K^*$, above which infinitesimal disorder is irrelevant. The latter is a
decreasing function of the width $n$. Moreover, the fluctuations enhance the Josephson
tunneling  between the ladders. Both of these effects lead to a higher $T_c$ for the array.

\section{The Single Ladder Model}

We begin by considering a single fluctuating $n$-leg ladder.
We assume that the fluctuations occur on length scales which
are long compared to the electronic wavelength and thus neglect backscattering
from them. Specifically, we treat the limit where the ladder oscillates rigidly
inside a parabolic potential. Its state is given by the deviation $u(\tau)$ from
the bottom of the well (taken in the following
as the $x$-axis), and its action in imaginary time is
\be
\label{Su}
S_u=\int d\tau\left[\frac{M}{2}(\partial_\tau u)^2+\frac{M\omega_0^2}{2}u^2 \right] ,
\ee
where $M$ is the ladder's effective mass. The fluctuations are characterized by the
oscillation frequency $\omega_0$, and by their typical amplitude $\lambda=1/\sqrt{M\omega_0}$.

The non-interacting spectrum of the ladder consists of $n$ energy bands. We assume
that the filling is such that all of them cross the Fermi level. If this is not the case,
then, in the following, $n$ should be taken as the number of bands that do cross $E_F$. Each
of those do so at two Fermi points $\pm k^{(b)}_F$, one containing left-moving
($r=-1$), and the other right-moving ($r=1$) particles.
The annihilation operator of an electron in band $b$, with chirality $r$ and
spin projection $\sigma=\pm1$ can be expressed using bosonization language as
\be
\psi_{b,r,\sigma}(x)=\frac{F_{b,r,\sigma}}{\sqrt{2\pi a}}e^{i r k^{(b)}_Fx
-\frac{i}{\sqrt{2}}\left[\theta^{(b)}_c-r\phi^{(b)}_c+\sigma\theta^{(b)}_s
-\sigma r\phi^{(b)}_s\right](x)},
\label{bosop}
\ee
where the Klein factors $F_{b,r,\sigma}$ ensure the correct anti-commutation between
the different particle species, and $a$ is a short distance cutoff of the order of $k_F^{-1}$.
Physically, $\phi^{(b)}_c$ and $\phi^{(b)}_s$ are, respectively, the
phases of the $2k^{(b)}_F$ charge and spin-density wave fluctuations in band $b$, while
$\theta^{(b)}_c$ is the superconducting phase of the band.

When the electronic interactions are turned on, one finds for generic densities away
from half filling that the system ends up in a phase where all the spin fields
$\phi^{(b)}_s$, and all the relative charge modes $\theta^{(a)}_c-\theta^{(b)}_c$,
acquire a gap \cite{RGladders,num-ladders}. Consequently, the pronounced
fluctuations of their conjugated fields $\theta^{(b)}_s$, and $\phi^{(a)}_c-\phi^{(b)}_c$,
cause their correlations to decay exponentially to zero.
In this so called C1S0 phase the only gapless
excitations are created by the total charge fields
\be
\phi_{c+}=\frac{1}{\sqrt{n}}\sum_{b=1}^n\phi^{(b)}_c,~~~~~
\theta_{c+}=\frac{1}{\sqrt{n}}\sum_{b=1}^n\theta^{(b)}_c.
\label{c+}
\ee
Note that if the interaction strength is larger than the band-structure gaps, the number
of occupied bands may change relative to the occupancy of the bare spectrum, and $n$ should
be adjusted correspondingly.

The plane containing the ladder is taken to include a weak random potential which
couples to the electrons, and through them to $u$. This choice is motivated by the
situation in the cuprates where the "ladders" are defined by the sites occupied by
holes, and as such interact with the disorder only indirectly via the electronic degrees
of freedom. Consequently, the forward, $V^\eta({\bf r})$, and backward
scattering, $V^\xi({\bf r})$, disorder components couple as
\be
H_{\rm dis}=\int dx \sum_{a,b,j=1}^n  \left[ h^\eta_{a,b}(x,y_j)
+ h^\xi_{a,b}(x,y_j)\right] +{\rm H.c.} ,
\label{hdis}
\ee
where $y_j=jd+u$, with $d$ the inter-leg distance, and
\ba
\label{heta}
\nonumber
h^{\eta,\xi}_{a,b}({\bf r})&=&\sum_{r,\sigma} e^{i r(k_{F_a}\mp k_{F_b})x}
V^{\eta,\xi}_{a,b,r}({\bf r}) \psi^\dagger_{a,r,\sigma}(x)\psi_{b,\pm r,\sigma}(x) \\
&\sim&\sum_{r,\sigma}\frac{V^{\eta,\xi}_{a,b,r}({\bf r})}{2\pi a}e^{-i\sqrt{\frac{\pi}{2}}
\chi_{\eta,\xi}(a,b,r,\sigma)(x)}.
\label{hxi}
\ea
The exponents in Eq. (\ref{hxi}) are given by
\ba
\nonumber
\!\!\!\!\!\!\chi_{\eta,\xi}(a,b,r,\sigma)&=&r(\phi^{(a)}_c \mp \phi^{(b)}_c)-
(\theta^{(a)}_c - \theta^{(b)}_c) \\
&+&r\sigma(\phi^{(a)}_s \mp \phi^{(b)}_s)-\sigma(\theta^{(a)}_s -
\theta^{(b)}_s).
\label{chi}
\ea
They vanish for the intra-band forward scattering processes and instead these
terms take the form
\be
h^\eta_{a,a}({\bf r})\sim V^\eta_{a,a}({\bf r})
\left(n_a +\frac{\sqrt{2}}{\pi}\partial_x\phi_c^{(a)}\right),
\label{hforward}
\ee
where $n_a$ is the average electronic density in band $a$.
Assuming a short-range Gaussian disorder, one finds that its components obey
$\overline{V^{\eta}_{a,b,r}(x,y_j)[V^{\eta}_{a',b',r'}
(x,y_{j'})]^*}=DB^{j,j'}_{a,b}[\delta_{r,r'}\delta_{a,a'}\delta_{b,b'}
+\delta_{r,-r'}\delta_{a,b'}\delta_{b,a'}]
\delta(x-x')\delta(y_j-y_{j'})$, and
$\overline{V^{\xi}_{a,b,r}(x,y_j)[V^{\xi}_{a',b',r'}
(x,y_{j'})]^*}=DB^{j,j'}_{a,b}\delta_{r,r'}[\delta_{a,a'}\delta_{b,b'}
+\delta_{a,b'}\delta_{b,a'}]
\delta(x-x')\delta(y_j-y_{j'})$.
Here, $D$ is the disorder
strength and $B_{a,b}^{j,j'}$ are numbers of order unity which depend on the
details of the band structure.

In the following we focus on the physics at energies of the order
of the inter-ladder couplings, which we assume lie well below the scale $\Delta$,
set by the gaps of the massive modes. Under such conditions the latter can be replaced
by their non-vanishing expectation values, thus yielding
\ba
\nonumber
\chi_{\eta,\xi}(a,b,r,\sigma)&=&r(\phi^{(a)}_c \mp \phi^{(b)}_c)-
\sigma(\theta^{(a)}_s - \theta^{(b)}_s) \\
&+& c_{\eta,\xi}(a,b,r,\sigma) ,
\label{simpchi}
\ea
where $c_{\eta,\xi}(a,b,r,\sigma)$ are constants. All
the $\chi_{\eta,\xi}$ contain fields conjugated to the massive modes, whose correlation
functions decay exponentially. As a result, except for the intra-band forward
scattering part (whose effects we discuss in the next section)
the bare disorder potential coupling,
Eq. (\ref{hdis}), is irrelevant in the RG sense. However, higher order terms,
generated in the course of the RG by integrating over the massive modes, may be
relevant \cite{orignac2l,effectiveD}. To obtain the structure of these terms, which
are the ones responsible for the degradation of the pairing correlations on the ladder,
we need to consider all possible linear combinations of the $\chi_{\eta,\xi}$ with
integer coefficients. The most relevant coupling is that which contains the gapless
field $\phi_{c+}$ with the smallest coefficient in front of it, and no gapped
fields. Eq. (\ref{simpchi}) indicates that this coupling is generically\cite{comment1}
generated by $\sum_{b=1}^n \chi_\xi(b,b,r,\sigma)$,
and takes the form 2$\sum_{b=1}^n\phi^{(b)}_c=2\sqrt{n}\phi_{c+}$. Consequently,
the scattering by the effective disorder is described by
\be
H_{\rm eff-dis}=\int dx \frac{V_{\rm eff}(x,u)}{2\pi\alpha}e^{i\sqrt{2 n}\phi_{c+}(x)}
+{\rm H.c.} ,
\label{hdiseff}
\ee
where $\alpha$ is of the order of the correlation length $v/\Delta$ of the gapped fields,
with $v$ being their typical velocity \cite{comment2}. The leading
contribution to the effective disorder satisfies
$\overline{V_{\rm eff}(x,y)V^*_{\rm eff}(x',y')}=D_{\rm eff}\delta(x-x')\delta(y-y')$,
with
\be
\label{Deff}
D_{\rm eff}\sim D(D/a\lambda\Delta^2)^{n-1}(\Delta/\omega_0)^{(n-1)/2}(\alpha/a)^{n+1}.
\ee

Hence, after averaging over the disorder and the ladder dynamics, as described by
Eq. (\ref{Su}), we are led to consider
the following action for the C1S0 phase of the $n$-leg ladder
\begin{widetext}
\ba
\nonumber
S&=&\frac{1}{\pi}\int dx d\tau \left[ -i\partial_x\theta_{c+}\partial_\tau\phi_{c+}
+\frac{v_{c+}K_{c+}}{2}(\partial_x\theta_{c+})^2
+\frac{v_{c+}}{2K_{c+}} (\partial_x\phi_{c+})^2\right] \\
&-&D_{\rm eff}\int \frac{dx d\tau d\tau'}{(2\pi\alpha)^2}F(\tau-\tau')
\cos \left\{\sqrt{2 n}\left[\phi_{c+}(x,\tau)-\phi_{c+}(x,\tau')\right]\right\},
\label{S}
\ea
where the fluctuations kernel is given by \cite{fluctuating}
\be
F(\tau-\tau')=\langle\delta[u(\tau)-u(\tau')]\rangle
=\frac{1}{\sqrt{2\pi}\lambda}\frac{1}{\sqrt{1-e^{-w_0|\tau-\tau'|}}}.
\label{F}
\ee
\end{widetext}

\section{RG Analysis}

The RG analysis of this action is most naturally formulated in terms of the following
dimensionless parameters
\ba
\nonumber
&&\!\!\!\!\!\!\!\!\!\!\!\!\!\!\!\!
{\cal D}=\frac{nD_{\rm eff}\alpha}{\sqrt{2\pi^3}\lambda v^2_{c+}},~~~~~
\varpi=\frac{\alpha\omega_0}{v_{c+}}, \\
&&\!\!\!\!\!\!\!\!\!\!\!\!\!\!\!\!
K=K_{c+}-\frac{\cal{D}}{4}\left(\frac{4}{n^2}+K^2_{c+}\right)\int_0^1 ds
\frac{1}{\sqrt{1-e^{-\varpi s}}}.
\label{params}
\ea
Note that the parameter $K$ is no longer a pure measure of the electronic interactions,
but is rather an admixture of the interactions and the disorder \cite{comment3}.
However, it is $K$ which controls the asymptotic decay of the various
correlation functions \cite{giamsc}. The resulting RG flow equations,
with respect to the running cutoff $\alpha(\ell)=\alpha e^\ell$ are
\ba
\label{RG}
\nonumber
&&\hspace{-1cm}\frac{dK}{d\ell}=-\frac{1}{2}\frac{\cal D}{\sqrt{1-e^{-\varpi}}}K^2,
~~~~~~\frac{d{\cal D}}{d\ell}=(3-nK){\cal D}, \\
&&\hspace{-1cm}\frac{d\varpi}{d\ell}=\varpi-\frac{4{\cal L}{\cal D}}{n}\frac{(\cosh\varpi-1)}
{(1-e^{-\varpi})^{3/2}}, \\
\nonumber
&&\hspace{-1cm}\frac{d\lambda}{d\ell}=-\frac{2{\cal L}{\cal D}}{n}
\frac{(1-\cosh\varpi+\varpi\sinh\varpi)}{\varpi(1-e^{-\varpi})^{3/2}}\lambda,
\ea
with the dimensionless ladder length ${\cal L}=L/2\pi\alpha$ scaling
as $d{\cal L}/d\ell=-{\cal L}$. The details of their derivation are presented in
Ref. \onlinecite{fluctuating}.

The RG equations (\ref{RG}) imply that infinitesimal disorder is irrelevant
for $K>K^*=3/n$, independently of the fluctuations. More important,
however, are the effects of fluctuations in the presence of finite disorder.
First, since ${\cal D}\propto \lambda^{-n}\omega_0^{-(n-1)/2}$, see Eqs.
(\ref{Deff}) and (\ref{params}), large amplitude and frequency
oscillations tend to decrease the initial strength of the effective disorder.
Secondly, Eq. (\ref{RG}) shows that the disorder is more efficient
in renormalizing $K$ for smaller oscillation frequencies, and that it drives $\varpi$
towards this regime. Due to both mechanisms,
stronger and faster fluctuations cause $K$
to flow more slowly to smaller values. Consequently, the ladder's superconducting
pairing correlations, which decay as $x^{-1/nK}$, are less degraded by the disorder.
As demonstrated below, these stronger pairing correlations translate into an increase
of $T_c$ for the Josephson-coupled array.

Few comments are appropriate at this point. First, note that formally the static ladder limit
is reached in our model by taking $\lambda\rightarrow 0$, $\omega_0\rightarrow\infty$,
in order to keep the ladder in its ground state inside the well.
Secondly, as mentioned in the previous section, the intra-band forward scattering
terms, Eq. (\ref{hforward}), survive in the gapped phase. They are relevant in the
RG sense: the dimensionless strengths of the parts which couple to the average
band-density and to its long-wavelength fluctuations scale as
$d{\cal D}_{f,1}/d\ell=3{\cal D}_{f,1}$ and
$d{\cal D}_{f,2}/d\ell={\cal D}_{f,2}$, respectively.
These processes can be shown not to affect the pairing correlations in the
static limit \cite{giamsc}, or more generally when $\omega_0\rightarrow\infty$.
However, for finite $\omega_0$ they induce a retarded attraction between the
electrons via exchange of ladder's oscillation
quanta. This attraction tends to renormalize $K$ towards
higher values and enhance pairing, thus reinforcing the effects outlined above.
The intra-band forward scattering processes do contribute to the localization
of the ladder since they provide additional channels for
its coupling to the disorder [beyond the one contained in Eq. (\ref{S})].
Consequently, in the last two flow equations of set (\ref{RG}) one should
substitute ${\cal D}\rightarrow{\cal D}+{\cal D}_{f,1}+{\cal D}_{f,2}$.
Due to its large scaling dimension ${\cal D}_{f,1}$ is particularly effective
in localizing the ladder.

Finally, one should also take into account the renormalization of $K$ at energy scales
which span the range between $E_F$ and $\Delta$, {\it i.e.}, during the generation of
the gaps which define the C1S0 phase. Perturbative RG analysis of the single chain problem
\cite{fluctuating,giamsc} indicates that $K$ renormalizes by an amount of the order of
$D/v_F\lambda\Delta$ during this stage. In the following we assume that $D\ll v_F\lambda\Delta$,
and ignore this "high energy" renormalization. Moreover, such a condition is necessary
in order to ensure that disorder does not destroy the gaps. Since the size of the gaps
is known to decrease rapidly with the number of legs
\cite{chakracvarty-sg,ourreview,steve-RMP},
the region of applicability of our treatment shrinks correspondingly with $n$.

\section{Coupled Ladders}

Next, we consider an array of coupled C1S0 $n$-leg ladders.
The arguments leading to Eq. (\ref{hdiseff}) imply that the
most relevant inter-ladder charge-density wave (CDW) coupling, ${\cal V}$, is the $2nk_F$
CDW coupling. Its scaling dimension, $2-nK$, makes it more relevant
than the Josephson tunneling only once $K<1/n$. Moreover, $h^\eta_{a,a}$, which
we neglected in this work, is known to suppress ${\cal V}$ exponentially \cite{giamsc}.
Inter-ladder $q\sim0$ couplings also promote ${\cal J}$ over ${\cal V}$\cite{CDW}.
Owing to these reasons and our interest in applications to the superconducting
cuprates, we assume that the Josephson tunneling between adjacent ladders dominates over
their CDW coupling. Hence we analyze the effect of
\be
H_J=-\int dx \sum_{\langle i,j \rangle}\sum_{a,b=1}^nJ_{ab}(u_i-u_j){\Delta^{i}_a}^\dagger
\Delta^{j}_b + {\rm H.c.} ,
\label{HJ}
\ee
with $\Delta^{i}_a=\psi^{i}_{a,+,\downarrow}\psi^{i}_{a,-,\uparrow}+
\psi^{i}_{a,-,\downarrow}\psi^{i}_{a,+,\uparrow}$ the singlet superconducting
order parameter in band $a$ of the $i$-th ladder. The Josephson
tunneling amplitude depends, exponentially, on the inter-ladder separation,
$J_{ab}\sim J_0e^{-|s+u_i-u_j|/\gamma}$, where $s$ is the mean spacing of the array
and $\gamma$ depends on the environment between the ladders\cite{stevenature}.
Assuming $\lambda\ll s$, in order to remain in the quasi-one-dimensional limit,
we obtain for $H_J$ in the C1S0 phase
\be
H_J=-\int dx \sum_i \frac{J e^{(u_i-u_{i+1})/\gamma}}{(2\pi\alpha)^2}
\cos\left[\sqrt{\frac{2}{n}}\left(\theta^{i+1}_{c+}-\theta^{i}_{c+}\right)\right],
\label{bosHJ}
\ee
where\cite{comment2} $J=8\sum_{a,b=1}^n J_{ab}(0)\langle\cos(\sqrt{2}\phi^{(a)}_s\rangle
\langle\cos(\sqrt{2}\phi^{(b)}_s\rangle$. Its contribution to the RG flow,
in terms of ${\cal J}=J/2\pi\sqrt{n} v_{c+}$, and derived assuming\cite{comexpand}
${\cal J},\lambda/\gamma\ll 1$, is
\ba
\label{JRG}
\nonumber
&&\hspace{-0.5cm}\frac{\partial K}{\partial\ell}={\cal J}^2\left[1+\frac{\lambda^2}{\gamma^2}
(1+e^{-\varpi})\right],~~~~
\frac{\partial {\cal J}}{\partial\ell}=\left(2-\frac{1}{nK}\right){\cal J},\\
&&\hspace{-0.5cm}\frac{\partial\varpi}{\partial\ell}=\varpi-2n\pi^2{\cal L}{\cal J}^2
\frac{\lambda^2}{\gamma^2}\left[1+I_0(\varpi)\right],\\
\nonumber
&&\hspace{-0.5cm}\frac{\partial\lambda}{\partial\ell}=n\pi^2{\cal L}{\cal J}^2
\frac{\lambda^3}{\gamma^2}\frac{1+I_0(\varpi)-\varpi I_1(\varpi)}{\varpi},
\ea
where $I_{0,1}(x)$ are modified Bessel functions.

By integrating Eqs. (\ref{RG}) and (\ref{JRG}) one can calculate the scale, $\ell^*$,
at which ${\cal J}$ grows to be of order 1, and obtain an estimate for the transition
temperature of the array $T_c\sim\Delta e^{-\ell^*}$.
Since fluctuations tend to increase $K$, via its renormalization by both
${\cal D}$ and ${\cal J}$, and since $K$ controls the scaling of ${\cal J}$,
$T_c$ likewise rises.
Another estimate for $T_c$ can be derived using the inter-ladder mean field theory
\cite{suzumura,orignac2l,meanfield}. By replacing $Je^{(u_i-u_j)/\gamma}\rightarrow
\tilde{J}=2J\langle e^{(u_i-u_j)/\gamma}\rangle=
2Je^{\lambda^2/2\gamma^2}$, and $\cos[\sqrt{2/n}(\theta^{i}_{c+}-
\theta^{j}_{c+})]\rightarrow \langle\cos(\sqrt{2/n}\theta_{c+})\rangle\cos(\sqrt{2/n}
\theta_{c+})$ in $H_J$, the problem turns into that of a single ladder
described by Eq. (\ref{S}) and coupled to an effective field
\be
H_J\rightarrow -\int dx \frac{W}{(2\pi\alpha)^2}\cos\left(\sqrt{2/n}\theta_{c+}\right),
\label{effF}
\ee
induced by the neighboring ladders and determined self-consistently from
$W=\tilde{J}\langle\cos(\sqrt{2/n}\theta_{c+})\rangle$. At $T_c$, when $W$ is vanishingly small,
this condition reads
\be
\frac{1}{\tilde{J}}=\frac{1}{(2\pi\alpha)^2}\int dx\int_0^{1/T_c} d\tau P(x,\tau),
\label{Tceq}
\ee
with  $P(x,\tau)=\langle \cos[\sqrt{2/n}\theta_{c+}(x,\tau)]
\cos[\sqrt{2/n}\theta_{c+}(0,0)]\rangle$ calculated
using the action (\ref{S}). Approximating the effect of temperature by a cutoff
length $l_T=v_{c+}/T$ beyond which all correlators decay exponentially to zero, and using
the scaling properties of $P$ , Eq. (\ref{Tceq}) becomes \cite{suzumura,orignac2l}
\be
\frac{1}{\tilde{J}}=\int_\alpha^{v_{c+}/T_c} \!\!\frac{R dR}{4\pi v_{c+}\alpha^2}
\exp\left[-\int_0^{\ln\left(R/\alpha\right)}\frac{d\ell}{nK(\ell)}\right],
\label{TcMF}
\ee
where $K(\ell)$ is calculated from Eq. (\ref{RG}).

\begin{figure}[t!!!]
\setlength{\unitlength}{1in}
\includegraphics[width=3.4in]{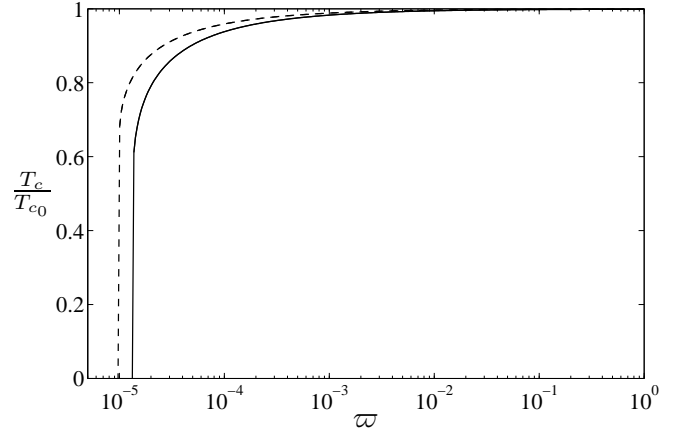}
\caption{$T_c$ of an array of 2-leg ladders as obtained by integrating the
RG equations (\ref{RG}) and (\ref{JRG}) (solid line), and from the mean field theory,
Eq. (\ref{TcMF}), (broken line). The results are for a system with bare $K=0.7$,
${\cal J}=10^{-2}$, ${\cal D}=10^{-4}$, $\lambda/\gamma=10^{-3}$, and ${\cal L}=10$.}
\label{Tcfig}
\end{figure}

Results for the ratio of $T_c$ to the transition temperature of the pure system, in
the case $n=2$, are presented in Fig. \ref{Tcfig}.
The effect of the fluctuations on $T_c$ is substantial\cite{noteTc} and is particularly
important whenever ${\cal D}$ is more relevant than ${\cal J}$,
{\it i.e.} for $K<(1+\sqrt{5})/2n$.
Fig. \ref{Tcfig} was generated for fixed bare ${\cal D}$, ignoring the dependence
of the effective disorder on $\omega_0$, as given by Eq. (\ref{Deff}). Taking this
dependence into account will only make the mitigating
consequences of the fluctuations more pronounced. Additional amplification of the physics
discussed here occurs when shape deformations of the ladders are taken into account
\cite{fluctuating}.

\section{Discussion}

Our study was motivated by the physics of the cuprate high-temperature superconductors,
especially the 214 compounds, which exhibit both static and fluctuating charge and spin
stripe orders over a substantial portion of their phase diagram\cite{ourreview,steve-RMP}.
In particular, we were interested to gain insight into the anomalous suppression of
$T_c$ at $x=1/8$, which is manifested to various degrees in these systems. One should
be cautious, however, when trying to apply the model studied here to the physical
materials. To begin with, we have considered the quasi-one-dimensional limit of
weakly coupled ladders. This limit allows for a controlled theoretical treatment
of the problem, but is an extreme caricature of the actual charge modulation
observed in the cuprates, whose amplitude is difficult to ascertain but which
is most likely weak. We then
assumed smooth long-wavelength fluctuations and ignored lattice commensurability
effects, which are believed to be important in the slowing down of the stripe
dynamics at $x=1/8$. In a related approximation we did not include fluctuations
of the stripe order towards a similar order oriented along the perpendicular
direction. Recent scanning tunneling spectroscopy\cite{kohsaka} demonstrated
domains of perpendicular stripe orientations in \BSCCO. It is conceivable,
however, that in the low temperature tetragonal phase of \LBCO, where the in-plane
rotational symmetry is broken and the $x=1/8$ anomaly is observed, such an approximation
is not too severe. Finally, the striped model possesses a large spin-gap, which implies
the absence of gapless nodal quasiparticles in the superconducting phase (although it
exhibits a "$d$-wave-like" order parameter). The same spin-gap means that the
model does not contain any of the low-energy incommensurate spin excitations,
which are the primary experimental evidence for stripes. These excitations originate
from the intervening lightly doped regions between the charged stripes.
However, apart from acting as effective barriers for tunneling between
the stripes [as characterized by the phenomenological length $\gamma$ in
Eq. (\ref{bosHJ})] they have little effect on the charge response of the system
governing its superconducting susceptibility, which is the focal point
of the present study.

Notwithstanding, our model constitutes a controlled theoretical laboratory for the
study of the interplay between the stripe dynamics and the disorder. Provided
that $\varpi$ is empirically associated with the distance from $x=1/8$
in the phase diagram of the lanthanum based cuprates, then Fig. \ref{Tcfig} offers a
mechanism for the $T_c$ anomaly at $x=1/8$. It suggests that the above mentioned
interplay plays an important role in alleviating, at least
partially, the disorder-induced suppression of superconductivity in these compounds.

\acknowledgments{
It is a pleasure to thank E. Arrigoni and T. Giamarchi for useful discussions. This
work was supported by the United States - Israel Binational Science Foundation
(grant No. 2004162) and by the Israel Science Foundation (grant No. 538/08).}


\begin{thebibliography}{999}

\bibitem{RGladders} L.~Balents and M.~P.~A.~Fisher, \prb {\bf 53}, 12133 (1996);
H.~J.~Schulz, {\it ibid.} {\bf 53}, R2959 (1996);
H.-H.~Lin, L.~Balents and M.~P.~A.~Fisher, {\it ibid.} {\bf 56}, 6569 (1997);
S.~Baruch and D.~Orgad, {\it ibid.} {\bf 71}, 184503 (2005).

\bibitem{num-ladders} R.~M.~Noack, S.~R.~White, and D.~J.~Scalapino, \prl {\bf 73},
882 (1994);
C.~A.~Hayward, D.~Poilblanc, R.~M.~Noack, D.~J.~Scalapino, and W.~Hanke,
{\it ibid.} {\bf 75}, 926 (1995);
S.~R.~White and D.~J.~Scalapino, \prb {\bf 57}, 3031 (1998);
S.~R.~White and D.~J.~Scalapino, {\it ibid.} {\bf 55}, R14701 (1997).
See also Refs. \onlinecite{ourreview,steve-RMP}.

\bibitem{manycomp} V.~J.~Emery, S.~A.~Kivelson, and O.~Zachar, {\prb } {\bf 59},
15641 (1999).

\bibitem{ourreview} E.~W.~Carlson, V.~J.~Emery, S.~A.~Kivelson, and D.~Orgad,
in {\it "Superconductivity: Novel Superconductors"}, Vol 2, p. 1225 , edited
by K. H. Bennemann and J. B. Ketterson (Springer-Verlag 2008).

\bibitem{steve-RMP} S.~A.~Kivelson, I.~P.~Bindloss, E.~Fradkin, V.~Oganesyan,
J.~M.~Tranquada, A.~Kapitulnik, and C.~Howald, Rev. Mod. Phys. {\bf 75}, 1201 (2003).

\bibitem{suzumura} Y.~Suzumura and T.~Giamarchi, J.~Phys.~Soc.~Jpn. {\bf 58}, 1748 (1989).

\bibitem{orignac2l} E.~Orignac and T.~Giamarchi, \prb {\bf 56}, 7167 (1997).

\bibitem{fluctuating} U.~London, T.~Giamarchi, and D.~Orgad, \prb {\bf 73}, 134201 (2006).

\bibitem{lanthanum} J.~M.~Tranquada, B.~J.~Sternlieb, J.~D.~Axe, Y.~Nakamura,
S.~Uchida, Nature (London) {\bf 375}, 561 (1995);
M.~Fujita, H.~Goka, K.~Yamada, and M.~Matsuda, \prl {\bf 88}, 167008 (2002);
M.~Fujita, H.~Goka, K.~Yamada, J.~M.~Tranquada, L.~P.~Regnault, \prb {\bf 70},
104517 (2004).

\bibitem{la-str-Tc} J.~M.~Tranquada,  J.~D.~Axe, N.~Ichikawa, A.~R.~Moodenbaugh,
Y.~Nakamura, and S.~Uchida, \prl {\bf 78}, 338 (1997);
N.~Ichikawa, S.~Uchida, J.~M.~Tranquada, T.~Niem$\ddot{\rm o}$ller, P.~M.~Gehring,
S.-H.~Lee, and J.~R.~Schneider, {\it ibid.} {\bf 85}, 1738 (2000)

\bibitem{LBCO1/8} A.~R.~Moodenbaugh, Y.~Xu, M.~Suenaga, T.~J.~Folkerts, and
R.~N.~Shelton, \prb {\bf 38}, 4596 (1988).

\bibitem{effectiveD} E.~Arrigoni, B.~Brendel, and W.~Hanke, \prl {\bf 79}, 2297 (1997);
E.~Arrigoni and S.~A.~Kivelson, \prb {\bf 68}, 180503(R) (2003).

\bibitem{comment1} In general it can also be constructed using other combinations.
For example, in the 2-leg ladder it follows from $\chi_\xi(1,2,r,\sigma)
+\chi_\xi(1,2,r,-\sigma)$.

\bibitem{comment2} The prefactors in Eqs. (\ref{hdiseff}) and (\ref{bosHJ}) contain
combinations of products of Klein factors. We assume that we work in the subspace
of states where their eigenvalue is 1.

\bibitem{comment3} Consequently $K$ and $K_{c+}$ differ in their RG flow. The latter
actually converges towards a seperatrix at $K_{c+}=2/n$. This can be seen by transforming
to $\tilde\phi=\sqrt{n/2}\phi_{c+}$ and $\tilde\theta=\sqrt{2/n}\theta_{c+}$, in terms of
which the action describes spinless fermions moving on a single disordered chain, with
$\tilde{K}=(n/2)K_{c+}$. For $K_{c+}=2/n$ the fermions are non-interacting and the
disorder can not change this fact. See Refs. \onlinecite{giamsc,fluctuating} and
I.~V.~Gornyi, A.~D.~Mirlin, and D.~G.~Polyakov, \prb {\bf 75}, 085421 (2007).

\bibitem{giamsc} T.~Giamarchi and H.~J.~Schulz, \prb {\bf 37}, 325 (1988).

\bibitem{chakracvarty-sg} S.~Chakravarty, \prl {\bf 77}, 4446 (1996).

\bibitem{CDW} V.~J.~Emery, E.~Fradkin, S.~A.~Kivelson, and T.~C.~Lubensky, \prl {\bf 85},
2160 (2000); A.~Vishwanath and D.~Carpentier, {\it ibid.} {\bf 86}, 676 (2001).

\bibitem{stevenature} S.~A.~Kivelson, E.~Fradkin, and V.~J.~Emery, Nature (London)
{\bf 393}, 550 (1998).

\bibitem{comexpand} The latter condition allows us to expand the exponential
dependence of ${\cal J}$ on $u_i-u_{i+1}$, see Eq. (\ref{HJ}).

\bibitem{meanfield} D.~J.~Scalapino, Y.~Imry, and P.~Pincus, \prb {\bf 11}, 2042 (1975);
E.~W.~Carlson, D.~Orgad, S.~A.~Kivelson, and V.~J.~ Emery, {\it ibid.} {\bf 62}, 3422
(2000);
E.~Arrigoni, E.~Fradkin, and S.~A.~Kivelson, {\it ibid.} {\bf 69}, 214519 (2004).

\bibitem{noteTc} The fact that $T_c\simeq T_{c_0}$ for large $\varpi$ in Fig. 1
is a result of the small bare ${\cal D/J}$ ratio used in the calculation.

\bibitem{kohsaka} Y.~Kohsaka, C.~Taylor, K.~Fujita, A.~Schmidt, C.~Lupien, T.~Hanaguri,
M.~Azuma, M.~Takano, H.~Eisaki, H.~Takagi, S.~Uchida, and J.~C.~Davis, Nature (London)
{\bf 315}, 1380 (2007).

\end{thebibliography}
\end{document}